\documentstyle[12pt,epsf]{article}

\makeatletter
\@addtoreset{equation}{section}

\makeatletter
\def\vF{{\mbox{\boldmath $F$}}}
\def\vB{{\mbox{\boldmath $b$}}}
\def\vp{{\mbox{\boldmath $\partial$}}}
\def\vx{{\mbox{\boldmath $x$}}}
\def\vH{{\mbox{\boldmath $H$}}}
\def\vE{{\mbox{\boldmath $E$}}}
\def\Tr{\mathop{\rm Tr}}
\def\c{{\cos\phi}}
\def\s{{\sin\phi}}
\def\bF{{\bar F}}
\def\bA{{\bar A}}
\def\bx{{\bar x}}
\def\bPhi{{\bar\Phi}}
\def\bz{{\bar z}}
\def\bp{{\bar\partial}}

\newcommand{\be}{\begin{equation}}
\newcommand{\ee}{\end{equation}}
\newcommand{\bea}{\begin{eqnarray}}
\newcommand{\eea}{\end{eqnarray}}
\newcommand{\Pf}{\mathop{\rm Pf}}
\newcommand{\nn}{\nonumber}
\newcommand{\vs}[1]{\vspace*{#1}}
\newcommand{\hs}[1]{\hspace*{#1}}
\newcommand{\p}{\partial}
\newcommand{\Half}{\frac12}
\newcommand{\unit}{\hbox to 3.8pt{\hskip1.3pt \vrule height 7.4pt
    width .4pt \hskip.7pt \vrule height 7.85pt width .4pt \kern-2.4pt 
    \hrulefill \kern-3pt \raise 3.7pt\hbox{\char'40}}}
\newcommand{\wbar}{\overline}
\newcommand{\del}{\partial}
\newcommand{\VEV}[1]{\left\langle #1 \right\rangle}

\begin{document}

\begin{titlepage}

\title{
\vspace{-10mm}
\hfill\parbox{4cm}{
{\normalsize KEK-TH-718}\\[-5mm]
{\normalsize KUNS-1687}\\[-5mm]
{\normalsize NFS-ITP-00-114}\\[-5mm]
{\normalsize\tt hep-th/0010026}
}
\\
\vspace{15mm}
Symmetry Origin of Nonlinear Monopole
}
\author{
{}
\\
Koji {\sc Hashimoto}${}^1$\thanks{{\tt koji@itp.ucsb.edu}},
\hs{1mm}
Takayuki 
{\sc Hirayama}${}^2$\thanks{{\tt thirayam@ccthmail.kek.jp}}
\hs{1mm}
and
\hs{1mm}
Sanefumi
{\sc Moriyama}${}^3$\thanks{{\tt moriyama@gauge.scphys.kyoto-u.ac.jp}}
\\[15pt]
${}^1$ {\it Institute for Theoretical Physics,}\\
{\it University of California, Santa Barbara, CA 93106}\\[7pt]
${}^2$ {\it High Energy Accelerator Research Organization (KEK),}\\
{\it Tsukuba, Ibaraki 305-0801, Japan}\\[7pt]
${}^3$ {\it Department of Physics, }\\
{\it Kyoto University, Kyoto 606-8502, Japan}\\[7pt]
}
\date{\normalsize October, 2000}
\maketitle
\thispagestyle{empty}

\begin{abstract}
\normalsize\noindent
We revisit the non-linear BPS equation: the Dirac monopole of the
Born-Infeld theory in the $B$-field background.
The rotation used in our previous papers to discuss the scalar field
by transforming the BPS equation into a linear one is extended to the 
case of gauge field.
We also find that this transformation is a symmetry of the action.
Moreover using the Legendre-dual formalism we present a simple
expression of the BPS equation.
\end{abstract}

\end{titlepage}

\section{Introduction}

In the recent progress in string theory, non-commutativity gets
attention as one of the crucial aspects of string theory. 
String theory by itself is a theory of non-local objects and therefore
should have a close relation with non-commutativity
\cite{WitSFT,STUP}.

Among plenty of situations with the non-commutativities, there is an
intriguing situation which has an alternative commutative description:
D-brane in NS-NS $B$-field background.
When the two-form $b$ is polarized along the worldvolume directions of
the D-brane, then the low energy effective physics on the D-brane is
described by two equivalent theories.
One is non-commutative Dirac-Born-Infeld (DBI) theory with
$*$-product \cite{CDS,DH}, and the other is ordinary DBI theory in the
backgournd $B$-field \cite{ACNY}.
These two are related with each other by field redefinitions
\cite{SW}.

Using this interesting equivalence, solitons in the non-commutative
theories have been studied 
\cite{NS}--\cite{GN2}.
This is partly because the non-commutativity produces new types of
solitons which do not exist in theories without non-commutativities,
such as $U(1)$ instantons \cite{NS}.
Another reason is that in the non-commutative gauge theory the
monopole solution has a non-local structure \cite{HasHas,HHM,GN} which
is one of the key properties of the string theory behind it.
The solitons can be interpreted in terms of the equivalent ordinary
descriptions with $B$-field. However in general, the equations which
described the solitons become highly non-linear, hence called
{\it non-linear BPS equation}.

In our previous papers \cite{hh,moriyama}\footnote{The author in
  \cite{Mat} discussed the electric BIon case.}, we considered $U(1)$
monopoles in the 4 dimensional non-commutative gauge theory, or
equivalently the ordinary DBI theory with the $B$-field background.
Since two effective descriptions are related by the field redefinition
as explained above, it is natural to expect that the solitons in the
two theories should also be related by the field redefinition.
In fact it was checked \cite{hh,moriyama} up to the first few orders
in the non-commutativity parameter $\theta$.

 From the solution for the scalar field it is possible to obtain a
physical picture in the commutative side using the brane
interpretation \cite{callan}.
Although a naive condition of preserving the linearly realized
supersymmetry of the DBI theory gives a picture of a tilted D3-brane
solution with a perpendicular D-string (see Fig.\ (i))
\cite{koji}\footnote{This is a persuasive result because the
  $B$-field serves as the magnetic field and the D-string slanted due
  to the force balance between the magnetic force and the string
  tension \cite{koji}.},
this is not the case for our non-linear BPS equation.
Instead, the non-linear BPS equation gives us a slightly different
picture with the horizontal D3-brane (Fig.\ (ii)) since the derivative
of the scalar field should vanish asymptotically by definition.
Hence, from these asymptotic behavior it was conjectured \cite{hh}
that these two pictures (i) and (ii) are related by the rotation in
the target space.
The exact manipulation was first performed in \cite{moriyama} and the
exact solution for the scalar field is obtained as a result.
Recently the result is lifted to M-theory \cite{Mic,You} by
considering M2-brane ending on M5-brane with a constant $C$-field
background.

\begin{figure}[tbp]
\begin{center}
\leavevmode
\epsfxsize=100mm
\epsfbox{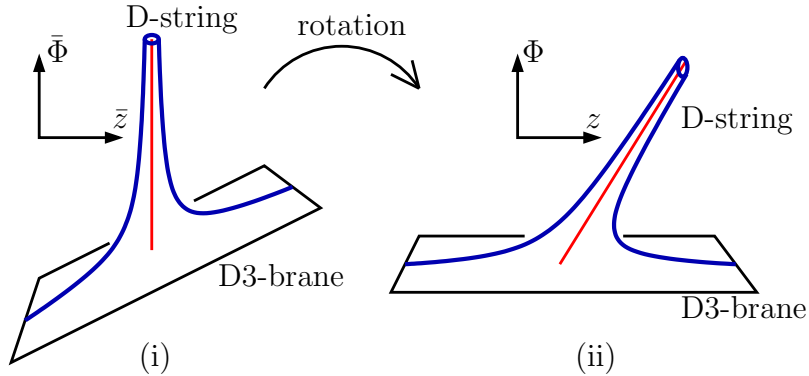}
\put(-66,90){$z$}
\put(-245,90){$\bar{z}$}
\put(-30,18){D3-brane}
\put(-175,125){rotation}
\put(-205,30){D3-brane}
\put(-270,115){$\bar{\Phi}$}
\put(-90,115){$\Phi$}
\put(-240,128){D-string}
\put(-30,90){D-string}
\put(-70,0){(ii)}
\put(-235,0){(i)}
\caption{Rotation in the target space from (i) to (ii). }
\label{fig:rot}
\end{center}
\end{figure}

However, in our previous papers the following questions remains to be
answered: What is the rotation?
How does it act on the gauge field?
Is the rotation a symmetry of the supersymmetrized DBI theory?
How is the relation between this rotation and linearly or non-linearly
realized supersymmetries?

In this paper, we shall answer these questions.
In addition to the above questions on the rotation transformation in
the target space,
we shall also study the BPS equation in the dual form to make the
symmetry of the BPS equation clearer.
Speculation for the generalization of our argument to the case of
instantons and non-abelian solitons shall be also given.

This paper is organized as follows. First in Sec.\ 2, we shall review
the non-linear BPS equations for monopoles, and show how the explicit
solutions of the non-linear BPS equations are obtained by taking the
rotation from the solutions of the linear BPS equations.   
In Sec.\ 3 we shall discuss the symmetry of the DBI action and find
that the rotation transformation we use to transform the non-linear
BPS equation into a linear one is a symmetry.
In Sec.\ \ref{sec:legen}, we present a simple form of the non-linear
BPS equations by taking the Legendre-dual of the magnetic field.
Final section is devoted to conclusions and discussions. We present
another derivation of non-linear BPS equation in App.\ A, and
App.\ B includes a simple form of the non-linear BPS equation for
instantons. 

\section{Non-linear Monopole}
\label{sec:nlmonopole}

In this section, we shall consider the non-linear monopole equation.
The monopole solution is a static solution in 3+1 dimensions and
a stable one because it satisfies the BPS condition\footnote{In this
article we loosely use the word BPS equation. Although originally the 
BPS equation is defined as a condition of satisfying the Bogomol'nyi
bound, in this paper we use it as the condition of preserving half of
the supersymmetries because the two conditions usually coincide.}.
Since now we turn on the NS-NS 2-form background, the BPS equation is
deformed into a non-linear one.
This monopole can be interpreted as a D-string ending on the D3-brane
\cite{callan}.
Due to the magnetic charge the D-string has on its end,
the D-string is tilted \cite{HasHas}.

Hereafter we shall solve the non-linear BPS equation by transforming
it into a linear one.
This transformation was conjectured in \cite{hh} and exact
manipulation was first achieved in \cite{moriyama}.
In the previous work \cite{moriyama} one of the authors discussed the
transformation by eliminating the gauge field because we did not know
how to treat it then. However here we shall also incorporate the gauge
field.

\subsection{Scalar field}
In this subsection we shall revisit the scalar field part of the
non-linear BPS equation \cite{moriyama}.
First we shall recall how the non-linear BPS equation is
obtained\footnote{Another derivation of the non-linear BPS equation
is presented in App.\ A.}.
For this purpose we have to consider the supersymmetrized
DBI action.
The linearly realized supersymmetries $\delta_{\rm L}$ and
non-linearly realized supersymmetries $\delta_{\rm NL}$ of the gaugino 
$\lambda$ are given as \cite{bag,ket,tse,SW}
\bea
\delta_{\rm L}\lambda
&=&\frac{1}{2\pi\alpha'}M_{mn}^+\sigma^{mn}\eta,\label{L}\\
\delta_{\rm L}\bar\lambda
&=&\frac{1}{2\pi\alpha'}M_{mn}^-\sigma^{mn}\bar\eta,\\
\delta_{\rm NL}\lambda&=&\frac{1}{4\pi\alpha'}
\Bigl(1-\Pf M+\sqrt{1-\Tr M^2/2+(\Pf M)^2}\Bigr)\eta^*,\label{NL}\\
\delta_{\rm NL}\bar\lambda&=&\frac{1}{4\pi\alpha'}
\Bigl(1+\Pf M+\sqrt{1-\Tr M^2/2+(\Pf M)^2}\Bigr)\bar\eta^*,
\eea
where $M$ denotes
\bea
M=(2\pi\alpha')\pmatrix{0&-\p_1\Phi&-\p_2\Phi&-\p_3\Phi\cr
\p_1\Phi&0&(F_3+b_3)&-(F_2+b_2)\cr
\p_2\Phi&-(F_3+b_3)&0&(F_1+b_1)\cr
\p_3\Phi&(F_2+b_2)&-(F_1+b_1)&0},
\eea
with the magnetic field $F_i\equiv\epsilon_{ijk}F_{jk}/2$, a constant
NS-NS 2-form background $b_i\equiv\epsilon_{ijk}b_{jk}/2$
($i,j,k=1,\ldots,3$) and the scalar field $\Phi$.
Here we turn on only the spatial components of the field strength,
the NS-NS 2-form and only one component of the scalar fields.
The matrix $M$ is obtained by the Wick rotation: We regard the
Euclidean time component of the gauge field as the scalar field $\Phi$
and discard the time derivatives.
We shall set $2\pi\alpha'=1$ for simplicity hereafter, however we can
restore it on the dimensional ground anytime we like.

We shall consider the monopole solution with the field strength $\vF$
and the derivative of the scalar field $\vp\Phi$ vanishing at the
infinity.
The non-linear BPS equation \cite{SW,marino} is the condition of
preserving the linear combination of $\delta_{\rm L}$ and
$\delta_{\rm NL}$ which is unbroken at the infinity:
\bea
\frac{\vF+\vB-\vp\Phi}{1+(\vF+\vB)\cdot\vp\Phi+
\sqrt{1+(\vF+\vB)^2+(\vp\Phi)^2+\Bigl((\vF+\vB)\cdot\vp\Phi\Bigr)^2}}
=\frac{\vB}{1+\sqrt{1+\vB^2}}.
\label{originalBPS}
\eea
Here the left hand side is the ratio between the linear (\ref{L}) and
non-linear (\ref{NL}) supersymmetries and the right hand side is its
value at the infinity.
The other combinations of supersymmetries are broken completely.
Note that the expression inside the square root of the left hand
side is rewritten as
\bea
1+(\vF+\vB)^2+(\vp\Phi)^2+\Bigl((\vF+\vB)\cdot\vp\Phi\Bigr)^2
=\Bigl(\vF+\vB-\vp\Phi\Bigr)^2
+\Bigl(1+(\vF+\vB)\cdot\vp\Phi\Bigr)^2,
\label{simplifyL}
\eea
which is the sum of square of the numerator $\vF+\vB-\vp\Phi$ and
square of the rest of the denominator $1+(\vF+\vB)\cdot\vp\Phi$.
This means that the BPS equation (\ref{originalBPS}) is simplified as
\bea
\frac{\vF+\vB-\vp\Phi}{1+(\vF+\vB)\cdot\vp\Phi}
=\frac{\vB}{1}.
\label{BPS}
\eea

The strategy adopted in \cite{moriyama} to map this non-linear BPS
equation (\ref{BPS}) into a linear one and solve it was to eliminate
the gauge field since we did not know how to incorporate it.
We shall first review the result in this subsection and see how to
take the gauge field into account in the next subsection.
First we note that eq.\ (\ref{BPS}) implies $\vF-\vp\Phi$ is
proportional to $\vB$:
\bea
\vF-\vp\Phi=f\vB,
\label{F-pPhi}
\eea
where $f$ is an unknown function.
By eliminating the field strength $\vF$ in eq.\ (\ref{BPS}), we find
\bea
f=(\vp\Phi)^2+(f+1)\vB\cdot\vp\Phi.
\label{f=Pf}
\eea
Another equation for $f$ and $\Phi$ besides (\ref{f=Pf}) is obtained
by taking the divergence of the relation (\ref{F-pPhi}) and using the
Bianchi identity $\vp\cdot\vF=0$,
\bea
-\vp^2\Phi=\vB\cdot\vp f.
\label{bianchi}
\eea
Now we have a system of differential equations (\ref{f=Pf}) and
(\ref{bianchi}) for the scalar quantities $f$ and $\Phi$.
After eliminating $f$ we find quite a non-linear equation for $\Phi$:
\bea
\vp^2\Phi\Bigl(1-\vB\cdot\vp\Phi\Bigr)^2
+2\vB\cdot\vp\vp\Phi\cdot\vp\Phi\Bigl(1-\vB\cdot\vp\Phi\Bigr)
+\vB\cdot\vp\vB\cdot\vp\Phi\Bigl(1+(\vp\Phi)^2\Bigr)=0.
\label{higgsBPS}
\eea
Hereafter we shall suppose the constant background $\vB$ is in the $z$ 
direction when considering a concrete situation and rewrite the
equation (\ref{higgsBPS}) in the cylindrical coordinate
$(\rho,\varphi,z)$ with $x=\rho\cos\varphi$ and $y=\rho\sin\varphi$,
\bea
&&(\p_\rho^2\Phi+\p_\rho\Phi/\rho+\p_z^2\Phi)(1-b\p_z\Phi)^2
+2b(\p_\rho\p_z\Phi\p_\rho\Phi+\p_z^2\Phi\p_z\Phi)(1-b\p_z\Phi)\nn\\
&&\hs{3cm}+b^2\p_z^2\Phi(1+(\p_\rho\Phi)^2+(\p_z\Phi)^2)=0.
\label{nonlinear}
\eea

The idea to deal with this desperately intricate differential equation
was to use the following coordinate transformation
\cite{Mat,hh,moriyama}, which mix the scalar field and the coordinate:
\bea
\pmatrix{\bPhi\cr\bz}
=\pmatrix{\cos\phi&\sin\phi\cr-\sin\phi&\cos\phi}\pmatrix{\Phi\cr z},
\label{rotation}
\eea
by an angle $\phi$ with $\tan\phi=b$.
If we change our variables into those with bars, we find our
non-linear differential equation reduces simply to the Laplace
equation:
\bea
\bp_\rho^2\bPhi+\bp_\rho\bPhi/\rho+\bp_z^2\bPhi=0.
\label{linear}
\eea
Here we have used the following useful formulas:
\bea
\p_z\Phi&=&\frac{\c\bp_z\bPhi-\s}{\c+\s\bp_z\bPhi},\nn\\
\p_\rho\Phi&=&\frac{\bp_\rho\bPhi}{\c+\s\bp_z\bPhi},\label{derivative1}
\eea
and
\bea
\p_z^2\Phi&=&\bp_z^2\bPhi/(\c+\s\bp_z\bPhi)^3,\nn\\
\p_z\p_\rho\Phi&=&\Bigl[\c\bp_z\bp_\rho\bPhi
+\s(\bp_z\bp_\rho\bPhi\bp_z\bPhi-\bp_z^2\bPhi\bp_\rho\bPhi)\Bigr]/
(\c+\s\bp_z\bPhi)^3,\label{derivative2}\\
\p_\rho^2\Phi&=&\Bigl[(\c)^2\bp_\rho^2\bPhi
+2\c\s\Bigl(-\bp_z\bp_\rho\bPhi\bp_\rho\bPhi
+\bp_\rho^2\bPhi\bp_z\bPhi\Bigr)\nn\\
&&+(\s)^2\Bigl(\bp_z^2\bPhi(\bp_\rho\bPhi)^2
-2\bp_z\bp_\rho\bPhi\bp_z\bPhi\bp_\rho\bPhi
+\bp_\rho^2\bPhi(\bp_z\bPhi)^2\Bigr)\Bigr]/(\c+\s\bp_z\bPhi)^3,\nn
\eea
which can be obtained from the coordinate transformation
(\ref{rotation}) and the chain rules.

After rewriting the intricate non-linear equation (\ref{nonlinear})
into a linear one (\ref{linear}), the solution of the non-linear BPS
equation is obtained as
\bea
\Bigl((1+\vB^2)\vx^2-(\vB\cdot\vx)^2
-2\vB\cdot\vx\Phi+\vB^2\Phi^2\Bigr)\Phi^2=q^2,
\label{scalarsol}
\eea
by transforming the solution with bars
$\bPhi=q/\sqrt{\rho^2+\bz^2}+b\bz$ into those without bars using
(\ref{rotation}).

\subsection{Gauge field}
This subsection is devoted to the gauge field of the non-linear BPS
equation.
The explicit expression of the magnetic field in terms of the scalar
field was obtained from the relations (\ref{F-pPhi}) and (\ref{f=Pf})
as
\bea
\vF=\vp\Phi+\frac{(\vp\Phi)^2+\vB\cdot\vp\Phi}{1-\vB\cdot\vp\Phi}\vB.
\label{magnetic}
\eea
To discuss the gauge field we have to be careful about the
transformation matrix for the coordinate transformation
(\ref{rotation}):
\bea
\bF_{ij}=\frac{\p x^k}{\p\bx^i}\frac{\p x^l}{\p\bx^j}F_{kl}.
\label{transformmat}
\eea
In this way we are led to the following result.
\bea
&&\bF_{\rho\varphi}=F_{\rho\varphi}
-\s\bp_\rho\bPhi F_{\varphi z}+\s\bp_\varphi\bPhi F_{z\rho},\nn\\
&&\bF_{\varphi z}=(\c+\s\bp_\varphi\bPhi)F_{\varphi z},\label{cov}\\
&&\bF_{z\varphi}=(\c+\s\bp_z\bPhi)F_{z\varphi}.\nn
\eea
Here the coefficients of the field strength come from the
transformation matrix.
Rewriting the right hand side of eqs.\ (\ref{cov}) into fields with
bars by using (\ref{magnetic}) and (\ref{derivative1}), we find
finally
\bea
\bF_{\rho\varphi}=\bp_z\bPhi-b,&\bF_{\varphi z}=\bp_\rho\bPhi,&
\bF_{z\rho}=0.
\label{linearBPS}
\eea
This is exactly the condition of preserving the linear
supersymmetries (\ref{L}).
If we call the differential equation (\ref{nonlinear}) after
eliminating the field strength $\vF$ the equation of motion of the
scalar field $\Phi$, what we did in the previous subsection was to
transform the equation of motion into a linear one (\ref{linear}).
However in the present subsection we transform the BPS equation
(\ref{magnetic}) into a linear one (\ref{linearBPS}) directly.

As a byproduct of transforming the non-linear BPS equation (\ref{BPS}) 
into a linear one (\ref{linearBPS}) including the gauge field by
taking the transformation matrix into account, the gauge field part of 
the monopole solution is also obtained exactly. It is given by
\bea
A_i=\frac{\p\bx^k}{\p x^i}\bA_k,
\eea
where $\bA$ is the usual gauge field of the Dirac monopole as we found 
by now.
Hence the gauge field solution of the monopole equation
(\ref{magnetic}) in the cylindrical coordinate is given as
\bea
A_\varphi=\bA_\varphi=\frac{-q\bar\rho^2}{\sqrt{\bar\rho^2+\bar z^2}
\Bigl(\sqrt{\bar\rho^2+\bar z^2}+\bar z\Bigr)},
\label{gaugesol}
\eea
with $\bar\rho=\rho$ and $\bar z=\c z-\s\Phi$ and the other components 
of the gauge field vanishing.
As pointed out in \cite{moriyama}, the solution for the scalar field
$\Phi$ (\ref{scalarsol}) is multi-valued.
Hence, since the solution for the gauge field (\ref{gaugesol}) is
written in terms of the scalar field $\Phi$, the gauge field is also
multi-valued depending on which branch of the scalar field we are
discussing.

Therefore we conclude that all we have to do to incorporate the gauge
field is to take the transformation matrix into account.
Here we have completed our projects of finding the exact solution of
the non-linear BPS equation for monopoles as well as for instantons
\cite{moriyama2}.

\section{Symmetries of DBI Action}

In the previous section we have revisited the Dirac monopole under the
$B$-field background which is equivalent to the Dirac monopole on the
non-commutative spacetime.
The Dirac monopole with the scalar field vanishing asymptotically
satisfies the non-linear BPS equation.
After the transformation (\ref{rotation}) mixing the scalar field and
the coordinate, the non-linear BPS equation (\ref{BPS}) was mapped to
a linear one (\ref{linearBPS}).
The key point to discuss the field strength is to consider the
transformation matrix (\ref{transformmat}).

This kind of transformation which mixes the scalar field and the world 
volume coordinate should be unfamiliar to the readers.
Besides, the reason why the transformation matrix is necessary has
also to be clarified.
To answer these questions we need one principle.
The principle is that the transformation which is a symmetry of the
action is allowed.
Then all we have to do is searching for a symmetry which is compatible
with the transformation (\ref{rotation}).

In this section we shall discuss the symmetry of the DBI action and
the transformation rules for the fields before and after taking the
static gauge.
There are two types of symmetries.
One is local symmetries in the world volume:
the diffeomorphism invariance for world volume coordinates.
The other is global symmetries in the target spacetime:
Lorentz and translational symmetries in the target space.
The next two subsections are devoted to these two types of symmetries
respectively.
Although the content in this section might be already familiar,
we shall
discuss it at length because it plays an important role to answer the
question we pose in the previous paragraph.
When we discuss the BPS equations, the supersymmetry also plays a
crucial role.
We shall reserve the subsequent subsection for discussing the
supersymmetrized DBI action.

To fix the notation we shall review the DBI action shortly here.
The bosonic part of the D-brane action is 
\begin{eqnarray}
S=\int d^{p+1}x\;{\cal L}=T_p\int d^{p+1}x 
\sqrt{-\det\left(g_{\mu\nu}+F_{\mu\nu}+b_{\mu\nu}\right)},
\label{original}
\end{eqnarray}
where $T_p$ is the brane tension and we omit the dilation dependence
and the Chern-Simons terms.
The induced metric $g$ and the induced B-field $b$ are defined as
\begin{eqnarray}
g_{\mu\nu}&=&G_{MN}(X)\del_{\mu}X^M(x)\del_{\nu}X^N(x),\nn\\
b_{\mu\nu}&=&B_{MN}(X)\del_{\mu}X^M(x)\del_{\nu}X^N(x),
\end{eqnarray}
where $x^\mu$ $(\mu,\nu=0,1,\cdots,p)$ denote the world volume
coordinates and $X^M$ $(M,N=0,1,\cdots,$ $9)$ are the 10 dimensional
target space coordinates.
$G_{MN}$ and $B_{MN}$ are the target space metric and NS-NS 2-form
$B$-field.

\subsection{Symmetry in the target space}
When we discuss the gravity theory in the target space, the metric and
the NS-NS $B$-field should transform as
\bea
G'_{MN}&=&\frac{\p X^K}{\p X'^M}\frac{\p X^L}{\p X'^N}G_{KL},\nn\\
B'_{MN}&=&\frac{\p X^K}{\p X'^M}\frac{\p X^L}{\p X'^N}B_{KL},
\eea
for the local diffeomorphism $X'^M=X'^M(X)$.
If we fix the background, we are interested in the global symmetry
which preserves the background.
Especially, for the flat metric only the translation and the
Lorentz transformation in the target spacetime remain as symmetries.
The transformation rule for the translation is simple:
\begin{eqnarray}
X^M&\rightarrow&X'^M=X^M+\epsilon^M,
\end{eqnarray}
where $\epsilon^M$ are transformation parameters.
On the other hand, the transformation rule for the Lorentz
transformation is
\begin{eqnarray}
X^M&\rightarrow&X'^M=R^M_{\hs{1ex}N}X^N,
\end{eqnarray}
where $R^M_{\hs{1ex}N}$ is an element of the $SO(1,9)$ Lorentz group.

\subsection{Symmetry in the world volume}
The action (\ref{original}) has the local symmetry which we usually
call the diffeomorphism invariance:
\begin{eqnarray}
x^{\mu}&\rightarrow&x'^{\mu}=x'^{\mu}(x).
\end{eqnarray}
Under this transformation, fields $G_{MN}(X)$ and $B_{MN}(X)$ are
intact because they are fields of the target space.
$X^M(x)$ is transformed as a scalar while the derivatives
$\del_{\mu}$ and the gauge field $A_{\mu}$ are transformed as a
vector:
\begin{eqnarray}
\del_{\mu}&\rightarrow&\del'_{\mu}
=J^{\hs{.8ex}\rho}_{\mu}\del_{\rho},\nn\\
A_{\mu}(x)&\rightarrow&A'_{\mu}(x')
=J^{\hs{.8ex}\rho}_{\mu}A_{\rho}(x),
\end{eqnarray}
with $J^{\hs{.8ex}\rho}_{\mu}=\del x^{\rho}/\del x'^{\mu}$.
Therefore under the diffeomorphism the square root of the
determinant in the action is altered by the factor $\det J$, but this
factor is exactly canceled by the factor $\det J^{-1}$ which comes
from the jacobian of the integrand $d^4x$. Hence, the action is
invariant under the diffeomorphism.

We usually fix this local symmetry by taking the static gauge
$X^{\mu}(x)=x^{\mu}$
for all the directions of the world volume $\mu=0,\cdots,p$.

\subsection{Symmetry under static gauge}
The transformation rules for the action (\ref{original}) are simple,
however, after taking the static gauge they become complicated.
It is because these transformation rules usually break the static
gauge and to retain the static gauge we must compensate the breaking
using the diffeomorphism invariance.

To retain the static gauge,
{\it i.e.} $\delta X^{\mu}=0 \; (\mu=0,\cdots,p)$,
for the translation, we must compensate the breaking by using the
following diffeomorphism invariance,
\begin{eqnarray}
x'^{\mu}=x^{\mu}+\epsilon^{\mu}.
\end{eqnarray}
By combining the translation symmetry in the target space and
diffeomorphism invariance, we find
\begin{eqnarray}
X^{\mu}(x)(=x^{\mu})&\rightarrow&
X'^{\mu}(x')= X^\mu(x)+\epsilon^{\mu}(=x'^{\mu}),\nn\\
X^i(x)&\rightarrow&X'^i(x')=X^i(x)+\epsilon^i,\\
A_{\mu}(x)&\rightarrow&A'_{\mu}(x')
=\frac{\del x^{\rho}}{\del x'^{\mu}}A_{\rho}(x)=A_{\mu}(x).\nn
\end{eqnarray}
We can easily check that under this transformation the breaking of the 
static gauge is restored:
\bea
\delta X^{\mu}(x)\equiv X'^{\mu}(x)-X^{\mu}(x)
=\Bigl(X^{\mu}(x-\epsilon)+\epsilon^{\mu}\Bigr)-X^{\mu}(x)
=(x^{\mu}-\epsilon^{\mu})+\epsilon^{\mu}-x^{\mu}=0. 
\eea 
It is apparent that the action is invariant under this transformation,
since we only use some combination of transformation in the target
spacetime and the diffeomorphism.
Since $\VEV{\delta_{\epsilon}X^i}\neq 0$, $X^i$ is the Nambu-Goldstone
(NG) boson for the spontaneous breaking of the translation for the
$i$-direction.

Next we consider Lorentz transformation. As before to retain the
static gauge, we must compensate the breaking by using the following
diffeomorphism invariance,
\begin{eqnarray}
x'^{\mu}=R^{\mu}_{\hs{.8ex}\nu}x^{\nu}+R^{\mu}_{\hs{.8ex}i}X^i(x),
\end{eqnarray}
and so the field transformations in the static gauge are
\begin{eqnarray}
X^{\mu}(x)(=x^{\mu})&\rightarrow&
X'^{\mu}(x')=R^{\mu}_{\hs{.8ex}\nu}X^{\mu}(x)
+R^{\mu}_{\hs{.8ex}i}X^i(x)(=x'^{\mu}),\nn\\
X^i(x)&\rightarrow&
X'^i(x')=R^i_{\hs{.8ex}\mu}X^{\mu}(x)+R^i_{\hs{.8ex}j}X^j(x) 
=R^i_{\hs{.8ex}\mu}x^{\mu}+R^i_{\hs{.8ex}j}X^j(x),\label{ro}\\
A_{\mu}(x)&\rightarrow&
A'_{\mu}(x')=\frac{\del x^{\rho}}{\del x'^{\mu}}A_{\rho}(x)
=\Bigl((R^{-1})^{\hs{.8ex}\rho}_{\mu}
+(R^{-1})^{\hs{.8ex}\rho}_j\del'_{\mu}X'^j(x')\Bigr)A_{\rho}(x).\nn
\end{eqnarray}
Even when $B$-field is zero, $\VEV{\delta X^i}$ is generally
non-zero.
It is natural because the rotation which mixes transverse and normal
directions to the brane is broken.
Note that $\VEV{\delta X^i}$ is also non-zero for the translation.
Therefore the NG mode for the breaking of the translation is also the
NG mode for the broken rotation transformation, and thus 
no new NG mode appears \cite{sun1,sun2,kug}.

Though it is obvious that the action is invariant under this
rotation,
we shall check the invariance explicitly to see the transformation
(\ref{rotation}) and (\ref{transformmat}) we used in the previous
section to map the non-linear BPS equation to a linear one is exactly
a symmetry.
Therefore we take the situation as in the previous section.
The Lagrangian in the static gauge with the target space flat and
$B_{12}$ non-zero is
\bea
S=\int d^4x\sqrt{-\det(g_{\mu\nu}+F_{\mu\nu}+b_{\mu\nu})},
\label{dbistatic}
\eea
with $g_{\mu\nu}=\eta_{\mu\nu}+\del_{\mu}\Phi\del_{\nu}\Phi$ and
$b_{12}=B_{12}$.
Here we omit the brane tension and put the scalar fields zero except
only one mode denoted by $\Phi$.
Now we consider the rotation which mixes the $X^3=x^3$ and $\Phi$
with angle $\phi$:
\begin{equation}
R^M_{\hs{1ex}N}=\pmatrix{\c&\s\cr-\s&\c},
\end{equation}
for $X^3$ and $\Phi$ directions. This $R$ is an identity matrix for
other directions.
With this representation of $R$ and eqs. (\ref{ro}), fields are
transformed as follows:
\begin{eqnarray}
\pmatrix{\Phi(x)\cr z}&\rightarrow&\pmatrix{\Phi'(x')\cr z'}
=\pmatrix{\c&-\s\cr\s&\c}\pmatrix{\Phi(x)\cr z},\nn\\
(t,x,y)&\rightarrow&(t',x',y')=(t,x,y),\nn\\
g_{\mu\nu}(x)&\rightarrow&g'_{\mu\nu}(x')
=G'_{MN}\del'_{\mu}X'^M\del'_{\nu}X'^N
=J_{\mu}^{\hs{.8ex}\rho}J_{\nu}^{\hs{.8ex}\sigma}g_{\rho\sigma}(x),
\\
F_{\mu\nu}(x)&\rightarrow&F'_{\mu\nu}(x')
=J^{\hs{.8ex}\rho}_{\mu}J^{\hs{.8ex}\sigma}_{\nu}F_{\rho\sigma}(x),
\nn\\
b_{\mu\nu}(x)&\rightarrow&b'_{\mu\nu}(x') 
=J_{\mu}^{\hs{.8ex}\rho}J_{\nu}^{\hs{.8ex}\sigma}b_{\rho\sigma}(x).
\nn
\end{eqnarray}
Then the action (\ref{dbistatic}) is invariant since the change from
$d^4x$ is exactly canceled by the change from the square root of the
determinant and so under the transformation we consider in the previous
section the action is invariant.

\subsection{Supersymmetrized DBI action}
In Sec.\ 2 we have constructed the non-linear BPS equation by
preserving a combination of the linear supersymmetry $\delta_{\rm L}$
(\ref{L}) and non-linear supersymmetry $\delta_{\rm NL}$ (\ref{NL})
that is unbroken at the infinity.
Further we find that the non-linear BPS equation (\ref{BPS}) is mapped
to a linear condition (\ref{linearBPS}) by the rotation transformation 
(\ref{rotation}) which mixes the scalar field and one of the world
volume coordinate.
Therefore, it might seem that the rotation transformation mix the
linear supersymmetry $\delta_{\rm L}$ and the non-linear supersymmetry 
$\delta_{\rm NL}$.

In the previous subsections, we have considered the symmetry of the
DBI action.
However, to discuss the supersymmetry we have to consider the
supersymmetrized DBI action. Recall that 
there are two types of the superstring action: the NSR formalism
preserves the supersymmetry in the world sheet and the GS formalism
preserves the supersymmetry in the target space.
In the same way, there are two types of the supersymmetrized DBI
action describing the D-brane which have been known
\cite{bag,ber,aga,ket}.
One is that has supersymmetry in the worldvolume \cite{bag, ket} and
the other is that possesses supersymmetry in the target space
\cite{ber, aga}.
It is believed they are related by some field redefinition, but the
explicit form of the field redefinition 
has not been known.
In Sec.\ \ref{sec:nlmonopole} we have derived the non-linear BPS
equation using the action possessing supersymmetry in the world
volume.
However, as in App.\ A it is possible to derive the same non-linear
BPS equation from the action which has supersymmetry in the target
space.
This fact strongly supports the folklore that two actions describe the
same physics.

To discuss the BPS equation it is preferable to use the DBI action
which has supersymmetry in the world volume, because it is more
transparent whether the linear supersymmetry $\delta_{\rm L}$ and the
non-linear one $\delta_{\rm NL}$ mix.
This formalism is derived using the technique of the non-linear sigma
model for the Goldstino which appears from the partial breaking for
${\cal N}=4$ down to ${\cal N}=2$ supersymmetry.
However, since the Goldstone multiplet in its components is too
complicated, we shall discuss our question using the action which has
supersymmetry in the target space instead.

The action which has supersymmetry in the target space \cite{ber,aga}
with the flat target space is 
\bea
S=\int d^{p+1}x~ L
=T_p\int d^{p+1}x\sqrt{\det g_{\mu\nu}+{\cal F}_{\mu\nu}},
\label{susydbi}
\eea
where the metric $g_{\mu\nu}$ and the field strength
${\cal F}_{\mu\nu}$ are given as
\bea
g_{\mu\nu}&=&G_{MN}\Pi^M_{\mu}\Pi^N_{\mu},\nn\\
{\cal F}_{\mu\nu}&=&F_{\mu\nu}
-\del_{[\mu}X^M\bar\lambda\Gamma_M\del_{\nu]}\lambda,
\eea
with the fields pulled back by the super-covariant form
$\Pi^M_{\mu}=\del_{\mu}X^M-\wbar{\lambda}\Gamma^M\del_{\mu}\lambda$
instead of the bosonic form $\del_{\mu}X^M$.
Note that originally $\lambda$ is the fermionic coordinate of the
superspace in 10 dimensions, hence $\Gamma_\mu$ and $\Gamma_i$ are 10
dimensional gamma matrices and $\lambda$ is a Majorana-Weyl fermion in
10 dimensions even though $\lambda$ turns out to be the world volume
fermion\footnote{More concretely, we may independently introduce
  the local Lorentz frames for the curved target spacetime and for the
  curved world volume spacetime (they are identified after taking the
  static gauge) and introduce the vielbein. However since we consider
  the flat target spacetime and so the flat worldvolume metric in the
  following, we do not introduce the local Lorentz frames.}.
Note also that originally the action has two Majorana-Weyl fermions and
is invariant under a local fermionic transformation (called $\kappa$
symmetry) as well as the diffeomorphism invariance.
To fix the $\kappa$ symmetry we take one of the fermions zero.
This fixing breaks neither the translation nor the Lorentz
transformation in the target space.

In string theory the $B$-field appears as the combination
$F_{\mu\nu}+b_{\mu\nu}$.
Therefore, to include the constant $B$-field background we have only
to replace $F_{\mu\nu}$ by $F_{\mu\nu}+b_{\mu\nu}$.
Finally the supersymmetric action including the $B$-field after fixing
the  
diffeomorphism invariance as well as the $\kappa$ symmetry is
explicitly given by,
\begin{equation}
S=\int d^4x\sqrt{\det(\eta_{\mu\nu}+F_{\mu\nu}+b_{\mu\nu}
+\del_{\mu}X^i\del_{\nu}X^i
-2\wbar{\lambda}(\Gamma_{\mu}+\Gamma_i\del_{\mu}X^i)\del_{\nu}\lambda
+\wbar{\lambda}\Gamma^M\del_{\mu}\lambda
\wbar{\lambda}\Gamma_M\del_{\nu}\lambda}.
\end{equation}
Here we do not regard the metric and the $B$-field as fields but only
as a background. Therefore the dynamical fields are only the gauge
field $A_{\mu}$, the fermion $\lambda$ and the scalar fields $X^i$.

The supertransformation in this theory is
\begin{eqnarray}
\delta\wbar{\lambda}&=\!\!&\wbar{\epsilon}_1+\wbar{\epsilon}_2\zeta 
+\xi^{\mu}\del_{\mu}\wbar{\lambda},\label{dlambda}\\
\delta X^i&=\!\!&(\wbar{\epsilon}_1-\wbar{\epsilon}_2\zeta)
\Gamma^i\lambda+\xi^{\mu}\del_{\mu}X^i,\label{dXi}\\
\delta A_{\mu}&=\!\!&(\wbar{\epsilon}_2\zeta\!-\!\wbar{\epsilon}_1)
(\Gamma_{\mu}+\Gamma_i\del_{\mu}X^i)\lambda
+\left(\frac{\wbar{\epsilon}_1}{3}\!-\!\wbar{\epsilon}_2\zeta\right)
\Gamma_M\lambda
\wbar{\lambda}\Gamma^M\del_{\mu}\lambda+\xi^{\rho}\del_{\rho}A_{\mu}
+\del_{\mu}\xi^{\rho}A_{\rho},\label{dAmu}
\end{eqnarray}
where
\bea
\xi^{\mu}=(\wbar{\epsilon}_2\zeta-\wbar{\epsilon}_1)
\Gamma^{\mu}\lambda,&&
\zeta=\frac{1}{\sqrt{\det M}}\frac{1}{(p+1)!}
\epsilon^{\mu_1\cdots\mu_{p+1}}\zeta_{\mu_1\cdots\mu_{p+1}},
\eea
with the infinitesimal parameters $\epsilon_1$ and $\epsilon_2$ which
are linear combinations of the linear supersymmetry and non-linear
supersymmetry.
Furthermore, if we take $p=3$, the number of supersymmetry is 
${\cal N}=4$
and $\zeta_{\mu_1\cdots\mu_{p+1}}$ is defined as
\bea
\zeta_{\mu_1\cdots\mu_4}=
\frac{1}{24}\gamma_{\mu_1}\cdots\gamma_{\mu_4}
+\frac{1}{2}{\cal F}_{\mu_1\mu_2}\gamma_{\mu_3}\gamma_{\mu_4}
+\frac{1}{2}{\cal F}_{\mu_1\mu_2}{\cal F}_{\mu_3\mu_4},
\eea
with $\gamma_{\mu}=\Gamma_M\Pi^M_{\mu}$ and
${\cal F}_{\mu\nu}=F_{\mu\nu}+b_{\mu\nu}
-\wbar{\lambda}(\Gamma_{\mu}+\Gamma_i\p_{\mu}X^i)\p_{\nu}\lambda
+\wbar{\lambda}(\Gamma_{\nu}+\Gamma_i\p_{\nu}X^i)\p_{\mu}\lambda$.

Now let us turn to our question posed at the beginning of this
subsection.
As said before, the fermion $\lambda$ is originally the fermionic
coordinates in the superspace of the target space and its indices
belong to the spinor representation of Lorentz group.
Therefore the transformation rule of the fermion is already decided.
Under the translation the fermion does not vary:
\begin{eqnarray}
\lambda'(x')=\lambda(x),&&
(x'^{\mu}=x^{\mu}+\epsilon^{\mu}),
\end{eqnarray}
and under the rotation fermion is transformed as a spinor:
\begin{eqnarray}
\lambda'(x')_{\alpha}=r_{\alpha}^{\hs{.8ex}\beta}\lambda(x)_{\beta},
&&
(x'^{\mu}=R^{\mu}_{\hs{.8ex}\nu}x^{\nu}+R^{\mu}_{\hs{.8ex}i}X^i(x)).
\end{eqnarray}
Here we denote $r$ as the spinor representation of $R$
($\in SO(1,9)$).
If we covariantly transform the parameter of the transformation, we
can find easily that the expression for the supersymmetry of the
gaugino $\lambda$ (\ref{dlambda}) and the gauge field $A_\mu$
(\ref{dAmu}) is not changed under the Lorentz transformation.
Since the Lorentz invariance is broken if we fix the diffeomorphism
invariance, the expression for the supersymmetry of the scalar field
(\ref{dXi}) apparently seems to be changed.
However, if we treat it carefully, it is found that terms proportional
to $(\wbar{\epsilon}_1-\wbar{\epsilon}_2\zeta)\Gamma^\mu\lambda$
disappear exactly and the expression of the supertransformation
remains invariant. 

Hence, from the above discussion on the DBI action which has
supersymmetry in the target space, we find that although the
non-linear BPS equation is mapped to a linear condition by the
rotation transformation, it does not mean that the linear
supersymmetry and the non-linear supersymmetry are mixed by the
rotation transformation. 
Instead, the rotation transformation changes the asymptotic behavior
of the fields and therefore maps the non-linear BPS equation into a
linear one.

Finally let us summarize this section.
It was surprising that the non-linear BPS equation is mapped to a
linear condition under the rotation transformation and the
transformation matrix of the field strength considered in the previous
section.
We have seen in this section that this transformation is allowed under
the principle that only the symmetry transformation of the action are
adopted. 
Finally we make it clear that our transformation does not mix the
linear supersymmetry and the non-linear one.
It just mixes the asymptotic behavior of the fields.

\section{Legendre-dual Formalism}
\label{sec:legen}

As is well known, the BPS equations of gauge theories are usually a
sort of electric-magnetic self-dual conditions.
However, this duality cannot be seen as a local symmetry of an action
in the usual Lagrangian formalism.
Nevertheless, if we restrict ourselves to static configurations, we
can make the duality explicit in the Legendre-dual formalism
\cite{Gibbons}.
Since in the spatial 3 dimensions the dual of a vector field is a
scalar field, the magnetic flux density is mapped to a magneto-static
potential under the Legendre transformation.
The Legendre-dual Lagrangian is manifestly symmetric under the
exchange of electro- and magneto-static potentials.

In our present situation, we have a scalar field $\Phi$, instead of
the electric field. But these two plays a similar role in the DBI
action \cite{Gibbons}, so we shall refer to the exchange between the
scalar field $\Phi$ and the magneto-static potential as duality in the 
following.

Since the non-linear BPS equation is a deformed self-duality
equation, it is natural to expect that this equation can be
understood also in the Legendre-dual formalism.
In this section we shall present an alternative formulation of the
non-linear BPS equation.
The rotation we used in Sec.\ 2 to map the non-linear BPS equation
into a linear one will be reformulated in a simple manner.
Here we will find a simple expression for the non-linear BPS
equation.
This simple form helps us to grasp the meaning of the non-linear BPS
equation.

First, we shall present in the Legendre-dual formalism the Lagrangian
we discussed previously to manifest the electric-magnetic duality
explicitly.
The self-duality condition will be identified with the condition of
preserving the linear supersymmetry.
After the warming up we shall proceed to study the non-linear BPS
equation in this formalism and find the rotation we considered in the
previous sections acts on the equation in a simple way.

\subsection{Legendre transformation and duality}
The Lagrangian under consideration is
\begin{eqnarray}
L=\sqrt{-\det(\eta_{\mu\nu}+\p_\mu\Phi\p_\nu\Phi
+F_{\mu\nu}+b_{\mu\nu})}.
\end{eqnarray}
Ignoring the electric field and restricting the configurations to
static ones, this Lagrangian can be written in the vector field
notation as
\begin{eqnarray}
L=\sqrt{1+|\vF+\vB|^2+|\vp \Phi|^2+(\vp\Phi\cdot(\vF+\vB))^2},
\end{eqnarray}
where the constant $B$-field $\vB$ is defined as 
$b_i=\epsilon_{ijk}b_{jk}/2$ as previously.
Now let us rewrite this action in the Legendre-dual formalism.
The magnetic flux density $\vF$ is subject to the Bianchi identity
$\vp\cdot\vF=0$, so if we consider the following Lagrangian
\begin{eqnarray}
\hat{L}\equiv L-\chi\vp\cdot\vF 
\end{eqnarray}
with the Lagrange multiplier scalar field $\chi$, then we can regard
$\vF $ as an independent variable. This is the Legendre 
transformation, and therefore the new Lagrangian $\hat{L}$ can be cast
into the function of $\chi$ and $\Phi$. 
After eliminating the field strength $\vF$ using the equations of
motion for $\vF$,
\begin{eqnarray}
\vp\chi=-\frac{\delta}{\delta\vF}\int\! d^3x\; L
=\frac{-1}{L}
\Biggl(\vF+\vB+[\vp\Phi\cdot(\vF+\vB)]\vp\Phi\Biggr),
\label{eq:defchi}
\end{eqnarray}
we obtain a Legendre-dual form
\begin{eqnarray}
\hat{L}=\sqrt{1+|\vp\Phi|^2-|\vp\chi|^2
-|\vp\chi|^2|\vp\Phi|^2+(\vp\chi\cdot\vp\Phi)^2}
-\vB\cdot\vp\chi.
\label{eq:legelag}
\end{eqnarray}
The last term is a surface term which of course does not affect the
equations of motion of this system, but we keep the term for later
discussion.
Note that in the above expression the Lagrangian depends on the
$B$-field only through the definition of $\chi$ (\ref{eq:defchi}).

Having discussed the Lagrangian, now we shall proceed to the condition 
of preserving supersymmetries.
We shall first discuss the condition in a heuristic way from the
symmetry of the action and then carefully check it later on.
Note that the Lagrangian (\ref{eq:legelag}) has the manifest global
symmetry $SO(1,1)$ which acts on fields as\footnote{This symmetry is
  the well-known electric-magnetic duality $SO(2)$ if we have the
  electro-static potential instead of $\Phi$.}
\begin{eqnarray}
\pmatrix{\bar\chi\cr\bar\Phi}
=\pmatrix{\cosh\theta&\sinh\theta\cr\sinh\theta&\cosh\theta}
\pmatrix{\chi\cr\Phi}.
\end{eqnarray}
Therefore, if one has a classical solution of this theory as
$\chi=\chi(x)$ and $\Phi=\Phi(x)$, then we can generate a series of
new solutions using this symmetry.
However, only when the initial classical solution satisfies the
following {\it (anti-)self-duality condition} 
\begin{eqnarray}
\chi=\pm\Phi,
\label{eq:asdcon}
\end{eqnarray}
it is impossible to obtain any new solution using the symmetry 
$SO(1,1)$\footnote{This is because the configuration satisfying the
self-duality condition is received only a constant multiplication
under the $SO(1,1)$ transformation.}.
Since in the ordinary quantum mechanics the ground state is usually
the most symmetric state, we should expect that this condition
(\ref{eq:asdcon}) is a kind of BPS condition, because the BPS state is 
also the state of the lowest energy among those having the same
topological quantities.
In fact, although our concern is the Dirac monopole which possesses
infinite energy, we shall see later that the above expectation 
is true in the present case: the condition (\ref{eq:asdcon}) is
equivalent to the one preserving the linear supersymmetries
(\ref{linearBPS}).

Note that the multiplier field $\chi$ has a meaning of magneto-static
potential, because the definition of $\chi$ (\ref{eq:defchi})
is precisely that of the true magnetic field $\vH$
\begin{eqnarray}
\vH=-\frac{\delta}{\delta\vF}\int\! d^3x\; L.
\end{eqnarray}
Thus if we regard the scalar field $\Phi$ as an electro-static
potential as explained in the introduction of this section, the
(anti-)self-duality condition becomes a familiar one: $\vH=\pm\vE$.
In the Maxwell electromagnetism, the Maxwell equations are invariant
under the interchange of $\vH\leftrightarrow\vE$ and
\mbox{\boldmath{$F$}} $\leftrightarrow$ \mbox{\boldmath{$D$}}.
Therefore the electric-magnetic duality is now explicitly realized.

As we have discussed, the (anti-)self-duality condition is a kind of
BPS condition. Therefore it is expected that this condition
consistently makes two equations of motion for two dynamical variables
reduce to a single equation. 
 From the Lagrangian $\hat{L}$ (\ref{eq:legelag}), we obtain two
equations of motion as
\begin{eqnarray}
&&\vp\cdot\left(\frac{1}{\hat{L}}
\Bigl(\vp\Phi-|\vp\chi|^2\vp\Phi+(\vp\Phi\cdot\vp\chi)\vp\chi\Bigr) 
\right)=0,\label{eq:1}\\
&&\vp\cdot\left(\frac{1}{\hat{L}}
\Bigl(-\vp\chi-|\vp\Phi|^2\vp\chi+(\vp\Phi\cdot\vp\chi)\vp\Phi\Bigr)
-\vB\right)=0.\label{eq:2}
\end{eqnarray}
In the present case $\vB$ is a constant vector and the last term does
not influence the equations of motion.
Hence, the two equations of motion are covariantly symmetric under the
duality group $SO(1,1)$.
Since the self-duality condition is a condition invariant under this
duality group, it is natural to expect the above reduction.
We can show explicitly that under the (anti-)self-duality
condition, eq.\ (\ref{eq:1}) is equivalent to eq.\ (\ref{eq:2}).

Furthermore, noting that $\hat{L}=1$ at the (anti-)self-dual point
(\ref{eq:asdcon}), we see that the equations of motion (\ref{eq:1})
and (\ref{eq:2}) reduce to a linear Laplace equation $\Delta\Phi=0$.
This equation is usually deduced from Maxwell-scalar theory.
Therefore this shows that the self-dual configuration has universal
footing among Maxwell theory and DBI theory \cite{callan,Gibbons}.
We have seen that the incorporation of the $B$-field does not affect
this property.

Finally we shall come back to show that the (anti-)self-duality
condition is equivalent to the condition of preserving the linear
supersymmetries (\ref{linearBPS})
\begin{eqnarray}
\vF+\vB=\mp\vp\Phi.
\label{eq:famil}
\end{eqnarray}
For this purpose, let us substitute the (anti-)self-duality condition
into the definition of $\vp\chi$ (\ref{eq:defchi}). Then we have 
\begin{eqnarray}
\vF+\vB+\Bigl(\vp\Phi\cdot(\vF+\vB)\pm L\Bigr)\vp\Phi=0.
\label{eq:const}
\end{eqnarray}
This indicates that $\vp\Phi$ is proportional to $\vF+\vB$. 
So if we put
\begin{eqnarray}
\vF+\vB=g\vp\Phi,
\end{eqnarray}
with some unknown scalar function $g$ and substitute this ansatz into
eq.\ (\ref{eq:const}), then we obtain $g=\mp 1$ explicitly.
Hence, we have shown that (\ref{eq:asdcon}) $\Rightarrow$
(\ref{eq:famil}).
The inverse, (\ref{eq:famil}) $\Rightarrow$ (\ref{eq:asdcon}), is also
true, as one can see easily from the definition of $\chi$
(\ref{eq:defchi}).

\subsection{Non-linear BPS equation and the rotation transformation}
So far we have discussed the condition of preserving the linear
supersymmetries, let us now proceed to interpreting the non-linear BPS
equation (\ref{BPS}) in the Legendre-transformed fashion. 
All we have to do is to substitute the equation (\ref{BPS}) into the
definition of $\vp\chi$ (\ref{eq:defchi}).
The result is surprisingly simple:
\begin{eqnarray}
-\sqrt{1+|\vB|^2}\;\vp\chi=\vp\Phi+\vB.
\label{eq:simple}
\end{eqnarray}
An unexpected fact is that the non-linear BPS equation is written in a
linear form, in spite of the complicated non-linear forms of both the
definition of $\chi$ (\ref{eq:defchi}) and the BPS equation
(\ref{BPS}).  

As in the previous subsection it should be important to see the
consistency with the equations of motion (\ref{eq:1}) and
(\ref{eq:2}).
Since under the condition (\ref{eq:simple}) we have a relation
\begin{eqnarray}
\lefteqn{\sqrt{1+|\vB|^2}\;
\left[\vp\Phi-|\vp\chi|^2\vp\Phi+(\vp\Phi\cdot\vp\chi)\vp\chi\right]}
\nn\\&&=
\left[-\vp\chi-|\vp\Phi|^2\vp\chi+(\vp\Phi\cdot\vp\chi)\vp\Phi\right]
-\hat{L}\vB,
\end{eqnarray}
it is straightforward to see that even the inside of the divergence in
eqs.\ (\ref{eq:1}) and (\ref{eq:2}) agree with each other.
Note that the surface term in proportion to a constant vector $\vB$
also appears consistently.
This is why we have kept it carefully.

As seen in the previous sections, it is natural to expect that
(\ref{eq:simple}) can be generated from the usual self-duality
condition $\chi=\pm\Phi$, by some transformation corresponding to the
rotation (\ref{rotation}).
In fact, we shall see this is the case in the following.
Let us consider the rotation which corresponds to the one
(\ref{rotation}) introduced in the previous sections:
\begin{eqnarray}
\pmatrix{\bar\Phi\cr\bar z}
=\pmatrix{\c&-\s\cr\s&\c}\pmatrix{\Phi\cr z},
\end{eqnarray}
and leave $x$, $y$ and $\chi$ intact.
Then the anti-self-duality condition $\chi=-\bar\Phi$ becomes
\begin{eqnarray}
-\chi=\Phi\;\c-z\;\s.
\label{eq:rotsd}
\end{eqnarray}
If we take $\tan\phi=b$ with $\vB=b\hat{\bf z}$ as in Sec.\ 2, this
condition (\ref{eq:rotsd}) is exactly the non-linear BPS condition
(\ref{eq:simple}).
In this derivation of the non-linear BPS condition using the
generalized rotation, the diffeomorphism, which was actually a bit
messy in Sec.\ \ref{sec:nlmonopole}, does not appear explicitly.
This comfortability is owing to the fact that all the entries in the
Legendre-transformed Lagrangian are scalar fields.

Before concluding this section, we point out that it is possible to
give a geometrical interpretation of the non-linear BPS condition.
Let us analyze the symmetry of the present system we are considering.
As we have noted, this system has the rotation symmetry $SO(4)$ and
the duality $SO(1,1)$.
These two apparently distinct symmetries are found to be combined to
form a larger rotation symmetry $SO(4,1)$.

Neglecting the ineffective surface term, the Lagrangian
(\ref{eq:legelag}) is cast into the form as 
\begin{eqnarray}
\hat{L}=\sqrt{\det(\delta_{ij}+\p_i\Phi\p_j\Phi-\p_i\chi\p_j\chi)},
\end{eqnarray}
with the indices $i,j=1,\ldots,3$.
This is a minimal surface action in the static gauge. We can easily
guess that the Lagrangian before the gauge fixing is written as
\begin{eqnarray}
\hat{L}=\sqrt{\det(\p_iu^A\p_ju^BG_{AB})},
\end{eqnarray}
where the metric $G_{AB}$ of the generalized target space is given by
\begin{eqnarray}
G_{AB}={\rm diag}(1,1,1,1,-1).
\end{eqnarray}
Because of the existence of the global rotation symmetry $SO(4,1)$
in the generalized target space, the gauge-fixed action has also the
symmetry $SO(4,1)$.
This is the generalization of the rotation (\ref{rotation}) considered
in the previous sections.
After the gauge-fixing, this symmetry has to accompany a local
diffeomorphism to attain the static gauge and becomes a symmetry
mixing the two fields $\chi$ and $\Phi$ and the world volume
coordinates.
Only the part which is not affected by the gauge fixing, original
$SO(1,1)$, is an intact global symmetry.

The self-duality corresponds to the null-surface in the 2
dimensional subspace ${\bf E}^{1,1}$ in the five dimensional
generalized target space ${\bf E}^{1,4}$.
Depending on the way of embedding of the null-surface of 
${\bf E}^{1,1}$ into ${\bf E}^{1,4}$, we have linear or non-linear BPS
equations.

In this section, we argued the non-linear BPS condition in the
Legendre-dual formalism, because the symmetry of the system is much
more manifest in the Legendre-dual form.
We further generalize the symmetry by combining the rotation group
$SO(4)$ and the duality group $SO(1,1)$.
Our analysis in this section not only makes the non-linear BPS
condition simple but also gives a geometrical understanding of the
condition.

An analogy of this understanding of the non-linear BPS equation may be
applied to the case of instantons.
This will be discussed in the final section, and we shall also present
a simple expression for the instanton nonlinear BPS equation in App.\
B.

\section{Conclusion and Discussion}

In this paper we analyze the gauge field part of the non-linear BPS
equation by transforming it into a linear one.
We complete our project of constructing the exact solution in the
Dirac-Born-Infeld theory.
The method is to incorporate the transformation matrix.
We also discuss the symmetry of the DBI action and find that the
rotation we used in transforming the non-linear BPS equation into a
linear one is indeed a symmetry.
Note also that the rotation does not mix the linear supersymmetry and
the non-linear one: it only changes the asymptotic behavior.
Another formulation of the BPS equation is also presented to discuss
the BPS equation in a much more compact form.

We expect our analysis in this paper has various applications.
We shall discuss a few future directions to conclude this paper.

First, since the rotation we used transforms the non-linear BPS
equation into a linear one beautifully for the monopole case, one
should expect a similar transformation is possible for the instanton
case.
We leave this question unanswered, because this generalization, even
if possible, seems not to be straightforward.
Furthermore, we can even present a negative argument for this
application.
In both the cases of monopole and instanton the effect of $B$-field in 
the condition of preserving the linear supersymmetry can always be
eliminated by a field redefinition which implies the moduli space is not
deformed by the $B$-field.
Therefore, as is pointed out in \cite{moriyama2} the transformation
that relates the non-linear BPS equation to a linear condition in the
monopole case is possible because the monopole moduli space is not
deformed by the non-commutativity parameter as can be seen from the
ADHM data \cite{Bak,GN}.
However, in the instanton case the small instanton singularity in the
moduli space disappears in general and the moduli space changes
drastically \cite{SW}.
Therefore 
if the moduli space can be defined in this case of singular
configurations, one does not expect a similar transformation exists
for the instanton case.
Though we present a formula for relating the field strength in
App.\ B, 
we find no local transformation for the gauge field.

Secondly, we would like to comment on the difficulty in the
generalization to the non-abelian case.
In the non-abelian case the BPS equation in the commutative side is
difficult because we do not know the precise form of the non-linearly
realized supersymmetry.
However we can still obtain the scalar field and the gauge field from
the solution of the non-commutative BPS equation using the field
redefinition of \cite{SW} in the expansion of the non-commutativity
parameter $\theta$.
After diagonalizing the scalar field and rotating the solution, it is
possible to find that the result agrees with the solution of the
linear condition to the first few orders \cite{hh,GH}.
However to generalize this argument to the gauge field is difficult
because even we diagonalize the scalar field, the gauge field is still
non-diagonal and therefore it is unclear how the transformation matrix
acts on the gauge field.
 From the viewpoint adopted in this paper, the transformation to be
performed should be a symmetry of the action.
But we find no similar symmetry in non-abelian DBI action
\cite{Tseytlin}. 

Finally we would like to comment on the zero slope limit.
It was pointed out in \cite{moriyama2} that the non-linear BPS
equation does not depend on the slope $\alpha'$ in the open string
moduli.
Therefore it should be natural to conjecture that the non-linear BPS
equation does not receive stringy derivative corrections.
In this paper we map the non-linear BPS equation into a linear one by
the rotation transformation.
The rotation transformation does not receive the stringy derivative
correction because it is a symmetry of the full string theory.
The linear BPS condition is also known \cite{Tho} not to receive the
correction.
Therefore this result strongly supports our conjecture that the
non-linear BPS equation does not receive the corrections.


\vs{10mm}

\noindent
{\Large\bf Acknowledgment}

We would like to thank T.\ Asakawa, Y.\ Michishita, T.\ Noguchi and
P.\ Townsend for valuable discussion.
This work is supported in part by Grant-in-Aid for Scientific
Research from Ministry of Education, Science, Sports and Culture of
Japan (\#02482, \#00843 and \#04633 for K.\ H.\ , T.\ H.\ and S.\ M.\
respectively), and by the Japan Society for the Promotion of Science
under the Postdoctoral/Predoctoral Research Program.
We are grateful also to the organizers of Summer Institute 2000 at 
Yamanashi, Japan, where a part of this work was carried out.

\appendix

\section{BPS Equation from Target Space Supersymmetries}

In Sec.\ 2, we have used the supersymmetrized DBI action which is
invariant under the worldvolume supertransformation.
The action is familiar as a non-linear generalization of
the super Maxwell action, and the monopole is a state which keeps a
half of the supersymmetries unbroken in that theory.
However, since the action in components is complicated, in Subsec.\
3.4, we have considered another action which is invariant under the
target space supertransformation instead.

It is believed that the two actions describe the same physics and they
are related by some field redefinition which is not known yet at
present.
However we can at least show that they give the same BPS equations.

Here we shall consider the supersymmetric Dirac monopoles and
instantons.
As shown in \cite{aga}, the supersymmetric DBI action is consistent
with the dimensional reduction by simply regarding the gauge fields in
the extra dimensions as scalar fields, so we have only to derive the
instanton BPS equation in the Euclidean frame.
We put fermions zero and excite no scalar field.
The Gamma matrices used in Subsec.\ 3.4 is replaced by the ones in
4 dimensions.
The BPS equation is derived from the condition that the vev of the
supertransformation for the fermions vanishes:
\begin{eqnarray}
\VEV{\delta\wbar{\lambda}}=\wbar{\epsilon}_1 
+\wbar{\epsilon}_2\frac{1}{\sqrt{\det M}}
\biggl(
\Gamma_5+(*{\cal F})_{\mu\nu}\Gamma^\mu\Gamma^\nu+\Pf{\cal F}
\biggr).
\end{eqnarray}
The Hodge dual operation $*$ is defined as
$(* F)_{\mu\nu}\equiv\epsilon_{\mu\nu\rho\sigma}F^{\rho\sigma}/2.$ 
Let us require that the field configuration vanishes at the infinity.
Then the surviving supersymmetries are a combination of the two
parameters,
\begin{eqnarray}
0=\wbar{\epsilon}_1+\wbar{\epsilon}_2 
\frac{1}{\sqrt{(1-\Pf b)^2+(b^+_{\mu\nu})^2}}
\biggl(
\Gamma_5+(*b)_{\mu\nu}\Gamma^\mu\Gamma^\nu+\Pf b
\biggr),
\label{eq:combi}
\end{eqnarray}
where we have used the following identity
\begin{eqnarray}
\det M=(1-\Pf{\cal F})^2+({\cal F}^+_{\mu\nu})^2
\end{eqnarray}
which is analogous to eq.\ (\ref{simplifyL}).
We have defined the (anti-)self-dual part of any matrix $A$ as
$A^\pm\equiv\Half(A\pm *A)$.

Now, we shall show that the following non-linear equation
\begin{eqnarray}
\frac{{\cal F}^+_{\rho\sigma}}{1-\Pf{\cal F}}
=\frac{b^+_{\rho\sigma}}{1-\Pf b}
\label{eq:nlbpsinst}
\end{eqnarray}
makes $\VEV{\delta\wbar{\lambda}} $ vanish for the combination
(\ref{eq:combi}) with restriction
\begin{eqnarray}
\wbar{\epsilon}_2\Gamma_5=\wbar{\epsilon}_2.
\label{eq:gamma5}
\end{eqnarray}
This is actually a straightforward calculation if one notes that 
\begin{eqnarray}
\sqrt{\det M}=
\left(1-\Pf{\cal F}\right)
\frac{\sqrt{(1-\Pf b)^2+(b^+_{\mu\nu})^2}}{\sqrt{(1-\Pf b)^2}},
\label{eq:LY}
\end{eqnarray}
and
\begin{eqnarray}
\wbar{\epsilon}_2(*{\cal F})_{\mu\nu}\Gamma^\mu\Gamma^\nu
=\wbar{\epsilon}_2{\cal F}^+_{\mu\nu}\Gamma^\mu\Gamma^\nu
\end{eqnarray}
which are valid under eqs. (\ref{eq:nlbpsinst}) and
(\ref{eq:gamma5}).

The equation (\ref{eq:nlbpsinst}) is in fact the non-linear BPS
equation originally derived from the DBI action supersymmetrized in
its world volume sense in \cite{moriyama2}. 
(See also \cite{SW, nlbps}.)
Therefore we have seen that eq.\ (\ref{eq:nlbpsinst}) is consistently
giving a BPS configuration also in the DBI theory supersymmetrized in
the target superspace method of Subsec.\ 3.4.
This fact is an evidence of the equivalence of the two formulations.

\section{Non-linear BPS Equation for Instanton }

In this appendix, we shall give a simple expression for the non-linear 
BPS equations for instantons, on the analogy of the non-linear
monopoles of Sec.\ \ref{sec:legen}. Since instantons are not static
configurations, we cannot apply the same method to obtain a scalar
field as a dual of the gauge field. However, we can introduce a dual
gauge field and its field strength using the usual Legendre-dual in
4 dimensions.
The dual field strength $P_{\mu\nu}$ is defined\footnote{
In our notation, when we take the differentiation, $F_{\mu\nu}$ should
be independent of $F_{\nu\mu}$. We are following the notation of 
\cite{gibhas}.
} as 
\begin{eqnarray}
P^{\mu\nu}\equiv -2\frac{\delta L}{\delta F_{\mu\nu}}.
\label{eq:defg}
\end{eqnarray}
As is easily seen, if the theory is Maxwell electrodynamics, then we
have $P=F$. 
Using this dual field strength, the non-linear BPS equation for
instantons will be able to be written in a form
\begin{eqnarray}
*P_{\mu\nu}=F_{\mu\nu}+b_{\mu\nu}
\end{eqnarray}
with some prefactors. We shall show this below.

\subsection{Self-duality of BI system in the $B$-field background}

First, we shall present a review of the duality invariance of the
4 dimensional Euclidean BI system \cite{gibhas, gibras}.
Especially we are incorporating the background $B$-field, but this
does not affect the ordinary argument as long as the linear BPS 
conditions are concerned. The self-dual constant $B$-field\footnote{In
  this appendix, we assume that the $B$-field is self-dual for
  simplicity.} is included in the Lagrangian as a combination
${\cal F}=F+b$. Using useful $SO(4)$ invariant variables
$X\equiv{\cal F}_{\mu\nu}{\cal F}^{\mu\nu}/4$ and
$Y\equiv\Pf{\cal F}={\cal F}_{\mu\nu}(*{\cal F})^{\mu\nu}/4$, 
the Lagrangian is given by
\begin{eqnarray}
L=\sqrt{\det(\delta_{\mu\nu}+{\cal F}_{\mu\nu})}=\sqrt{1+2X+Y^2}.
\label{eq:biaction}
\end{eqnarray}
Therefore we obtain the dual gauge field strength as 
\begin{eqnarray}
P_{\mu\nu}=-\frac{1}{L}
\left({\cal F}_{\mu\nu}+Y(* {\cal F})_{\mu\nu}\right).
\end{eqnarray}
Since the equations of motion and Bianchi identities are
$ \p^\mu P_{\mu\nu} = 0$ and $\p^\mu (* {\cal F})_{\mu\nu} =0$,
then the duality invariance of these equations are implemented as 
\begin{eqnarray}
\delta_{\rm d}{\cal F}=-*P,\quad\delta_{\rm d}P=-*{\cal F}.
\label{eq:dualrot}
\end{eqnarray}
In order for this symmetry to be valid, one has to require a
constitutive relation
\begin{eqnarray}
*{\cal F}=-\delta_{\rm d}P({\cal F})
=-\frac{\delta P}{\delta{\cal F}}\delta_{\rm d}{\cal F}
=\frac{\delta P}{\delta{\cal F}}*P.
\label{eq:constitu}
\end{eqnarray}
This equation together with the definition of $P$ (\ref{eq:defg})
gives a constraint on the form of the Lagrangian
\cite{const,gibras,gibhas}.  One can explicitly show that this
equation (\ref{eq:constitu}) is valid for the BI action
(\ref{eq:biaction}).

As seen from the equations of motion and the Bianchi identities, the
(anti-)self-duality condition reads 
\begin{eqnarray}
*P=\pm{\cal F}.
\label{eq:duality}
\end{eqnarray}
If we adopt the BI action, this is equivalent with the usual
(anti-)self-duality condition  $*{\cal F}=\pm{\cal F}$,
that is,
\begin{eqnarray}
{\cal F}^\mp=0.
\label{eq:bps}
\end{eqnarray}
This equivalence of (\ref{eq:duality}) and (\ref{eq:bps}) is one of
the nice properties of the BI theory \cite{gibhas}. 
And actually, it is possible to show explicitly that the action
(\ref{eq:biaction}) is bounded below by a topological quantity, and
the bound is given by this (anti-)self-dual configuration :
\begin{eqnarray}
S=\int\! d^4x\; (L-1)
=\int\! d^4x \sqrt{\left(1\pm Y\right)^2+2\left(X\mp Y\right)}-1
\geq\pm\int\! d^4x\; Y.
\end{eqnarray}
Here we have used semi-positive-definiteness of the quantity
\begin{eqnarray}
X\mp Y=\frac14\left({\cal F}_{\mu\nu}{\cal F}^{\mu\nu}
\mp{\cal F}_{\mu\nu}(*{\cal F})^{\mu\nu}\right)
=\frac18({\cal F}_{\mu\nu}\mp(*{\cal F})_{\mu\nu})^2
=\Half({\cal F^\mp}_{\mu\nu})^2
\end{eqnarray}
which vanishes when we adopt the (anti-)self-dual configuration.
Note that a natural self-duality condition is (\ref{eq:duality}), due
to the original duality transformation (\ref{eq:dualrot}).

\subsection{Non-linear BPS condition}
The condition (\ref{eq:bps}) is the so-called linear BPS condition. 
Assuming that the solution must be localized, then this linear BPS
condition requires an asymptotic behavior
\begin{eqnarray}
\lim_{r\rightarrow\infty}F_{\mu\nu}=-b^+_{\mu\nu}.
\end{eqnarray}
In \cite{SW,marino,nlbps,moriyama2} the non-linear BPS conditions
were considered. The solution of this condition has the asymptotic
behavior
\begin{eqnarray}
\lim_{r\rightarrow\infty}F_{\mu\nu}=0.
\end{eqnarray}
Therefore, so as to relate these two conditions, one can introduce a
parameter $u$ as
\begin{eqnarray}
\lim_{r\rightarrow\infty}F_{\mu\nu}=(u-1)b^+_{\mu\nu}.
\label{eq:u}
\end{eqnarray}
The value $u=1$ corresponds to the non-linear BPS condition, and $u=0$
corresponds to the linear BPS condition.

The combination of the linearly and non-linearly realized
supersymmetries which is consistent with the asymptotic behavior
(\ref{eq:u}) is
\begin{eqnarray}
\eta^*_\alpha=-u b_{\mu\nu}^+\sigma^{\mu\nu\;\beta}_{\;\;\alpha}
\eta_\beta.
\end{eqnarray}
Here $\eta$'s are parameters for the supersymmetry transformation, and
we adopted the notation of \cite{nlbps}.  
The BPS equation (which is not linear except $u=0$) is given by
\begin{eqnarray}
{\cal F}^+_{\mu\nu}=\Half ub_{\mu\nu}^+\left(1-Y+L\right).
\label{eq:nlbpsu}
\end{eqnarray}
This is a simple generalization of eq.\ (15) in \cite{nlbps}. 
Using the technique performed in the previous appendix, we can find
another equivalent expression of this equation as
\begin{eqnarray}
\frac{\cal F^+_{\mu\nu}}{1-Y}
=\frac{ub^+_{\mu\nu}}{1-\Pf(ub)}.
\label{eq:simesu}
\end{eqnarray}
This can be derived with the use of an equality that follows from eq.\
(\ref{eq:nlbpsu}) 
\begin{eqnarray}
\frac{1-\Pf(ub)}{1+\Pf(ub)}L=1-Y,
\label{eq:simesu2}
\end{eqnarray}
which is analogous to eq.\ (\ref{eq:LY}).
So one has to solve this non-linear BPS condition for obtaining
instanton configuration with the desired asymptotic behavior.

\subsection{Deformation of the self-duality condition}
Now, we want to show that a deformed (anti-)self-duality condition
\begin{eqnarray}
(*P)_{\mu\nu}=c_1{\cal F}_{\mu\nu}+c_2b_{\mu\nu}
\end{eqnarray}
with a certain choice of constant parameters $c_1(u)$ and $c_2(u)$
derives the non-linear BPS equation (\ref{eq:simesu}).
Taking the Hodge dual of the above deformed self-duality condition, we
have  
\begin{eqnarray}
(*P)^+_{\mu\nu}=c_1{\cal F}^+_{\mu\nu}+c_2b^+_{\mu\nu},
\quad(* P)^-_{\mu\nu}=c_1{\cal F}^-_{\mu\nu}.
\label{eq:plus}
\end{eqnarray}
Substituting the definition of $P$ into the second equation gives
a relation
\begin{eqnarray}
c_1L=1-Y.
\end{eqnarray}
Then using this relation and the first equation of (\ref{eq:plus})
gives  
\begin{eqnarray}
\frac{{\cal F}^+_{\mu\nu}}{1-Y}=\frac{-c_2b^+_{\mu\nu}}{2c_1}.
\end{eqnarray}
This functional form agrees with eq.\ (\ref{eq:simesu}) which we
wanted to see the equivalence. 

For the precise agreement, we have (see eqs.\ (\ref{eq:simesu}) and
(\ref{eq:simesu2}))
\begin{eqnarray}
c_1=\frac{1-\Pf(ub)}{1+\Pf(ub)},
\quad c_2=\frac{-2u}{1+\Pf(ub)}.
\end{eqnarray}
Thus the deformed self-duality condition which is equivalent with the
non-linear BPS condition is written as
\begin{eqnarray}
(*P)_{\mu\nu}=\frac{1-\Pf(ub)}{1+\Pf(ub)}{\cal F}_{\mu\nu}
+\frac{-2}{1+\Pf(ub)}ub_{\mu\nu}. 
\label{eq:final}
\end{eqnarray}
When $u=0$, evidently the above deformed self-duality condition
reproduces the linear BPS condition (\ref{eq:duality}). The
substitution $u=1$ gives a simple expression for the non-linear BPS
equation of instantons presented in \cite{marino,nlbps, moriyama2}.

We have rewritten the non-linear BPS equation in a simple form. One of
the virtue of this fact is that the consistency with the equation of
motion is manifest. Since the equation of motion is the vanishing of
the divergence of
$P_{\rho\sigma}$, using the above simple BPS equation one can easily
see that the equation of motion is reduced to the Bianchi identity. 
This fact is along the original motivation for using Bogomol'nyi
equations for finding explicit solutions.

\newcommand{\J}[4]{{\sl #1} {\bf #2} (#3) #4}
\newcommand{\andJ}[3]{{\bf #1} (#2) #3}
\newcommand{\AP}{Ann.\ Phys.\ (N.Y.)}
\newcommand{\MPL}{Mod.\ Phys.\ Lett.}
\newcommand{\NP}{Nucl.\ Phys.}
\newcommand{\PL}{Phys.\ Lett.}
\newcommand{\PR}{Phys.\ Rev.}
\newcommand{\PRL}{Phys.\ Rev.\ Lett.}
\newcommand{\PTP}{Prog.\ Theor.\ Phys.}
\newcommand{\hep}[1]{{\tt hep-th/{#1}}}
\newcommand{\hepp}[1]{{\tt hep-ph/{#1}}}

\end{document}